\documentclass[preprintnumbers,amsmath,amssymb,floatfix,twocolumn,prl]{revtex4}

\usepackage{dcolumn}
\usepackage{bm}
\usepackage{graphicx}

\begin{document}

\title {Fermi-surface geometry and ``Planckian dissipation''}
\author{ F. D. M. Haldane}
\affiliation{Department of Physics, Princeton University,
Princeton NJ 08544-0708}
\date{November  29, 2018}
\begin{abstract}
  A purely Fermi-surface formula is proposed for the Ohmic
  ``minimum metallic conductivity''
tensor  of clean metals with ``Planckian limit'' inelastic
dissipation.   This revises a recent proposal by Legros \textit{et al.}
\end{abstract}
\maketitle

Recently Legros \textit{et al.}\cite{legros}
have interpreted transport measurements on optimally-doped cuprates
(in the metallic regime that exhibit a $T$-linear resistivity above the
superconducting
transition) in terms of so-called ``Planckian dissipation''\cite{Zaanen,Bruin}
in which the relaxation time $\tau$ achieves a putative universal
minimum  value  $\hbar /k_BT$.
The authors of Ref.\cite{legros} propose a universal fit to the low-temperature metallic
Ohmic resistivity  of such materials (in zero magnetic field, with unbroken time-reversal
symmetry, so there is no Hall effect) of  a quasi-2D metal
of the form
\begin{equation}
 \rho(T)  = \rho_0    +  A_1dT
\end{equation}
where $d$  is the interlayer spacing, and   $A_1T$ is a ``universal''
(maximum) metallic resisitivity (of a single layer of a clean 2D metal
with only inelastic dissipation) given by
\begin{equation}
  A_1 T   =   \alpha (h/2e^2)(T/T_F)
    \label{formula}
\end{equation}
where $T_F$ is the ``Fermi temperature''  of the 2D electron gas,
defined as
\begin{equation}
  k_BT_F = \frac{\hbar^2k_F^2}{2m^*},
\end{equation}
where $m^*$ is an effective mass.

The formula (\ref{formula})
is reported in Ref.\cite{legros} to  be a good fit  to  measured results for a number of widely-differing
materials exhibiting $T$-linear resistivity,
with (to within about 30\% or so) $\alpha$ = 1.  
Taken at face value, this could suggest a  universal ``Planckian'' upper
bound  (with $\alpha$ =1)  on the strength of   dissipative inelastic scattering, so
that the conductivity
of a clean 2D metal satisfies
\begin{equation}
  \sigma(T) \ge \sigma_{\text{min}}(T)  = \left
    (\frac{2e^2}{h}\right )
\left (  \frac{\hbar^2k_F^2}{2m^*k_BT} \right ) .
\end{equation}

It is not the main  purpose here to explore the possible origin of
such a ``Planckian bound'', but instead  to suggest a more plausible
possible ``universal formula'', in the case that the bound is  valid.
An immediate objection to the proposed formula is to its use of the notional ``Fermi
temperature'', where $k_BT_F$ is the difference of  the
single-particle energy of an electron at the lowest point in the
conduction band, and that of an electron at the Fermi level.      The
experiments discussed in Ref.\cite{legros} are all carried out at
temperatures
$ T \ll T_F$, in which electrons at the bottom of the band (deep inside
the Fermi surface) cannot be
thermally excited, and, as a matter of principle, all low-temperature  conduction
processes in a metal  should involve only states at the Fermi level\cite{FDMH}.

With this in mind,  and setting $\alpha$ = 1,  the formula (\ref{formula}) for a
rotationally-symmetric 2D metal can be reinterpreted as
the conductivity tensor 
\begin{equation}
\sigma^{ab}_{\text{min}}(T) =
2\left (\frac{e^2}{\hbar}\right ) \frac{2\pi k_F}{(2\pi)^2} \left ( \frac{\hbar v_F}{k_BT}\right) \left (
  {\textstyle\frac{1}{2}}\delta^{ab}\right ).
 \label{revised}
\end{equation}
Here the initial factor of 2 counts the two  values of the spin
component of the  electrons,  $2\pi k_F$ is the arc-length  of
the 2D  Fermi surface in $k$-space,  $v_F$ is the Fermi velocity, and the final  isotropic
tensor factor is the geometric factor
\begin{equation}
\frac{1}{2\pi}\oint d\theta \, n^a(\theta) n^b(\theta)\equiv {\textstyle\frac{1}{2}}\delta^{ab},
\end{equation}
where $\bm n(\theta)$ = $(\cos \theta, \sin \theta)$ is the unit normal to
an isotropic Fermi surface, parametrized in the range $\theta \in [0,2\pi]$.
This is now a formula expressed entirely in terms of Fermi-surface
properties, and its generalization to arbitrary Fermi-surface
geometries can now be obtained.

The Fermi surface of a $d$-dimensional metal is a $(d-1)$-dimensional
oriented manifold characterized  by a Fermi vector $\bm k_F(\bm s)$,
and a Fermi velocity $\bm v_F(\bm s)$, which  points along the
outward normal to the Fermi surface in $k$-space, and defines its
orientation.  Here $\bm s$ = $\{s^1,\ldots , s^{d-1}\}$ is a
parametrization of the Fermi surface manifold.   There may be more
than one such manifold $\bm k_{F\alpha}(\bm s)$ labeled by a discrete
label $\alpha$.

While $k$-space is a flat Euclidean-like
$d$-dimensional space, it has no physically-preferred Euclidean metric.
Fermi-velocity components $\bm v_F$ = $v_F^a\bm e_a$ can be defined in terms
of a set of  constant basis vectors $\{\bm e_a, a=1,\ldots d\}$, where $\bm
e_a \cdot \bm e_b$ defines the symbol  ``$\delta_{ab}$'',
but the `` Euclidean metric'' $\delta_{ab}$
that this defines derives from an   arbitrary coordinate choice, and should not
enter in any physically-meaningful  formula.    The Fermi-velocity
components $v_F^a$ have upper (contravariant) indices, while the
components of the Fermi vector, given by $k_{Fa}$ = $\bm e_a\cdot \bm
k_{F\alpha}$, have lower (covariant) indices.     Only upper/lower
index-pairs can be contracted.

The Fermi-surface manifold $\bm k_{F\alpha}(\bm s)$
has a number of ambiguities.    First, under a constant
gauge-transformation
 $\bm k_{F\alpha}(\bm s)$ $\mapsto$ $\bm k_{F\alpha}(\bm s) - e\bm
A/\hbar$, so all physically-meaningful formulas must be invariant under 
a uniform shift in $k$-space.      Furthermore, $k$-space is
not an unbounded Euclidean space, but is its compactification into the
Brillouin zone, a $d$-torus with $d$-dimensional reciprocal (Bragg)
vectors $\bm G$.  The Fermi vector  of manifold $\alpha$ is ambiguous up to a shift
$\bm k_{F\alpha}$ $\mapsto$ $\bm k_{F\alpha}  + \bm G_{\alpha}$, where
$G_{\alpha}$ is any reciprocal vector.   If there are ``open orbits''
on the manifold, it supports non-trivial  closed paths $\Gamma$ where
\begin{equation}
  \oint_{\Gamma} \mathrm d \bm k_F \equiv \oint_{\Gamma} \frac{\partial \bm k_F}{ds} ds
  = \bm G_{\Gamma} \ne \bm 0,
\end{equation}
and (arbitrarily-chosen) Brillouin-zone boundaries, across which the
reduced $\bm k_F$ jumps by a  reciprocal vector, must be inscribed on
the manifold\cite{FDMH}.

Generically the Fermi surface manifolds are non-singular, with a non-vanishing
Fermi vector at each point, and are non-intersecting.    If both
time-reversal and spatial-inversion symmetry are unbroken, they have a
two-fold spin-degeneracy (to be included in the sum over labels
$\alpha$), otherwise spin-orbit coupling lifts the
degeneracy.   Pairs of Fermi surfaces can be degenerate at
isolated  points where a high-symmetry line in the Brillouin zone
intersects them, or on lines along which a high-symmetry plane
intersects them.  Other singularities such as  Weyl-point
degeneracies at generic points, or Van Hove singularities (where $\bm
v_F$ vanishes)  on the Fermi
surface, require fine-tuning in order to occur.

Only  the physically-interesting cases of dimensions $d$ = 1,2, and 3
will be discussed here.
Geometrically, one can first define the oriented  Fermi-surface
area $(d-1)$-forms (differential forms) $\bm  A_d$ = $A_d^a\bm
e_a$, which are (contravariant) vector forms.
With  labels distinguishing
distinct  manifolds  suppressed,
\begin{align}
   A_{d=1} &=  \xi  = \pm 1 ,    \\
 A^a_{d=2} &=  \epsilon^{ab}\mathrm dk_{Fb} \equiv  
                \epsilon^{ab}\frac{\partial k_{Fb}}{\partial s}ds,
                                 \\
  A_{d=3}^a &= \epsilon^{abc}\mathrm dk_{Fb}\wedge \mathrm dk_{Fc}
                 \equiv \epsilon^{abc}\frac{\partial k_{Fb}}{\partial
                 s^1}
                 \frac{\partial k_{Fc}}{\partial s^2}ds^1ds^2 .
  \end{align}
Here $\epsilon^{ab}$ and $\epsilon^{abc}$ are antisymmetric
Levi-Civita symbols.
As a vector, $\bm A_d$ is parallel to the Fermi velocity; this
defines a positive-scalar $(d-1)$-form $\tilde \gamma_d$:
\begin{equation}
  \bm A_d =   (2\pi)^d\bm v_F \tilde \gamma_d
\end{equation}
The universal $T$-linear low-temperature specific-heat of metals is
then given by
\begin{equation}
c = \frac{\pi^2}{3}k_B^2T\sum_{\alpha}\int_{\text{FS}_{\alpha}} \tilde
\gamma_d.
\label{tlinear}
\end{equation}

The Luttinger theorem that gives the band filling-factor $n_0$ (modulo
an integer) as a Fermi-surface integral is
\begin{equation}
  n_0 =
  \frac{1}{d}
  \frac{\Omega_d}{(2\pi)^d}\sum_{\alpha}\int_{\text{FS}_{\alpha}}
  \bm A_d\cdot \bm k_{F}
  \label{lutt}
\end{equation}
where $\Omega_d$ is the real-space volume of the ($d$-dimensional)
unit cell of the Bravais lattice, and
the ground-state momentum per unit cell is given  (modulo $\hbar$ times a reciprocal
vector) by
\begin{equation}
 \bm p_0 =
   \frac{\hbar }{d+1}
  \frac{\Omega_d}{(2\pi)^d}\sum_{\alpha}\int_{\text{FS}_{\alpha}}
  ( \bm A_d\cdot \bm k_{Fa})\bm k_F.
\end{equation}
These surface-integral expressions are valid for simple compact Fermi surfaces where
the point $\bm k = \bm 0$ is inside the surface.  Their ambiguities arise
when the Fermi surface in the Brillouin zone does not have  such a
simple topology, but their functional derivatives with respect to the
Fermi surface are unambiguous.

Note that
\begin{equation}
  \frac{\Omega_d}{(2\pi)^d}\sum_{\alpha}\int_{\text{FS}_{\alpha}}\
  \bm A_d   = 2\pi \bm a_{\alpha}.
\end{equation}
where $\bm a_{\alpha}$ is a primitive real space 
lattice translation, and  (if it is non-vanishing) is the ``Luttinger
anomaly'' of  a``chiral'' Fermi surface (such surfaces do not
individually
divide the
Brillouin zone into two regions, one ``inside'' and one ``outside'' the
oriented surface, so do not define an ``enclosed volume'' ).
Gauge invariance
requires that
\begin{equation}
  \sum_{\alpha} \bm a_{\alpha}  = 0
  \end{equation}
All Fermi points in $d$ =1 are chiral; the presence of chiral Fermi
surfaces in higher dimensions requires quasi-one-dimensionality: in
that case, the non-intersection property of Fermi-surfaces requires that
$\bm a_{\alpha}$ = $\xi_{\alpha} \bm a$ = $\pm \bm a$, where $\bm a$
is the lattice translation in the unique direction  (up to a sign) associated with
quasi-one-dimensionality.    For $ d > 1$, typical Fermi surfaces are
non-chiral, with $\bm a_{\alpha}$ = 0.

Now consider small fluctuations of the Fermi surface relative to its
ground state geometry:  $\bm k_{F\alpha}(\bm s) $ $\mapsto$
$\bm k_{F \alpha}(\bm s) + \delta \bm k_{F \alpha}(\bm s)$, and
define  the $(d-1)$-form $\rho$ = $\tilde \gamma_d\bm
v_F\cdot\delta\bm k_F$
This may be viewed as the net  density of quasiparticles minus
quasiholes associated with a patch (a $(d-1)$ form) of the Fermi surface.

To linear order in $\delta \bm k_F$,
the electric charge, electric  current,  and momentum
densities (relative to the ground state) are
\begin{equation}
 (J^0, \bm J, \bm \pi) =   \sum_{\alpha}\int _{\text{FS}_{\alpha}}
 \rho (e, e\bm v_F, \hbar \bm k_F).
 \label{excite}
 \end{equation}                         
 In contrast, the energy density relative to the ground state energy
 is  quadratic in $\delta \bm k_F$.

 In the presence of an applied electric field,
 \begin{equation}
   \frac{d\bm k_F}{dt} = \frac{e\bm E}{\hbar} ,
\end{equation}
or
\begin{equation}
  \frac{d \bm \rho(\bm s)}{dt}  =  (e/\hbar)\tilde \gamma_d(\bm s)  \bm v_F(\bm s)\cdot
  \bm E.
  \end{equation}  
Substitution into (\ref{excite}) confirms that,
\begin{equation}
  \frac{d}{dt} (J^0,  J^a, \pi_a) = (0, \Gamma^{ab} E_b,  (en_0/\Omega_d)E_a  ),
\end{equation}
where
\begin{equation}
 \Gamma^{ab} = \frac{e^2}{\hbar} \sum_{\alpha}
 \int_{\text{FS}_{\alpha}} \tilde \gamma _dv_F^av_F^b .
 \end{equation}

In the presence of an electric field, the electric current density
$\bm J$
grows linearly with time, without limit, in the absence of dissipation.
In a Drude-like phenomenology, dissipation is described by a
single quasiparticle lifetime $\tau$, and the time-evolution of the current density is
described by
\begin{equation}
  \left ( \frac{d}{dt} + \frac{1}{\tau}\right )  J^a = \Gamma^{ab}E_b.
  \label{drude}
 \end{equation}
 In particular, in a clean system with lattice translational symmetry,
 this (in the absence of Umklapp processes) allows the momentum that
 was added
 to the system by the action of the
 electric field to be absorbed by other unspecified non-electronic
 degrees of freedom by inelastic processes.     

If the suggestion of Ref.\cite{legros} that $\tau$ takes the proposed
universal limiting  minimum ``Planckian value'' $\hbar/k_BT$ in certain
situations is accepted, (\ref{drude}) leads to the
expression for this ``minimum metallic  conductivity'' at low
temperatures:
\begin{equation}
  \sigma_{\text{min}}^{ab}
=  \frac{e^2}{k_BT} \sum_{\alpha}\int_{\text{FS}_{\alpha}} \tilde
  \gamma_d v_F^av_F^b  .
\label{final}
\end{equation}
 For a simple rotationally-invariant 2D Fermi surface with two-fold
 spin degeneracy, this precisely agrees with the formula
  (\ref{revised}) given above as a reinterpretation
of the formula (\ref{formula}) proposed in  Ref.\cite{legros}. 
The conductivity tensor is related to the current-current
correlation function.   Since the expression (\ref{revised}) couples
only the velocities at the \textit{same} point of the Fermi surface, and is an
independent sum over patches, it can only describe an essentially-local
dissipation mechanism on the surface.

Note that the explicit appearance of Planck's constant has disappeared
in the formula (\ref{final}), but quantum mechanics is contained in
the Fermi-surface $(d-1)$-form $\tilde \gamma_d$ that controls the
single-particle density of states at the Fermi level, in both
(\ref{final}) and in  the formula (\ref{tlinear}) for the $T$-linear
specific heat.   The  density of one-electron quasiparticle
states per unit energy  at the Fermi level 
is explicitly given (per unit cell)  by
\begin{equation}
\mathcal D(E_F)  = \Omega_d \sum_{\alpha}\int_{\text{FS}_{\alpha}}
\tilde \gamma_d.
\end{equation}
This relation indicates that the positive scalar $(d-1)$-form  $\tilde
\gamma_d$ (which in less-mathematical notation is just
$dA/(2\pi)^d|v_F|$, where $dA$ is a Fermi surface scalar area element)
has a simple differential-geometry interpretation  as  
the natural  quantum  ``volume''   measure of the extent of
the 
$(d-1)$-dimensional
Fermi-surface manifold, while  $\bm A_d$ is the geometric $k$-space
area $(d-1)$-form that gives its $d$-dimensional  Luttinger-theorem
volume through  (\ref{lutt}).

The examples of $T$-linear resistivity discussed in Ref.\cite{legros}
all appear to involve systems near a  $T$ = 0 quantum critical point.
At such critical  points, spatial conformal invariance (and, if the
dynamical critical exponent is $z$ = 1, space-time conformal-invariance) is
an emergent property.   A key property of such conformal invariance is the
emergence of a unimodular (determinant 1)
spatial metric that parametrizes the conformal group.   It is not often emphasized that in an anisotropic crystalline
material without discrete microscopic rotational symmetries, this is a
truly emergent metric that has nothing to do with the Euclidean  metric of
empty space (just as the Lorentz velocity of emergent space-time
conformal invariance in $z$=1 quantum criticality
has nothing to do with the velocity of light in
empty space).

The relaxation of the momentum generated by the
action of the electric field on the charged electrons is the key
ingredient of the dissipation process that gives rise to steady-state
Ohmic conduction.     The divergent long-wavelength order-parameter
fluctuations associated with quantum criticality may enable a
Brownian-like diffusion of quasiparticles on the Fermi surface, moving
in a sequence of small jumps in $k$-space (near forward-scattering),
by absorbing or emitting
long-wavelength quanta of the  thermalized  critical-fluctuation field.

Such a mechanism may lead to a relaxation time that is uniform over
the Fermi surface, in contrast to dissipation described by a quantum
Boltzmann equation.   It is compatible with the evident locality of
the expression (\ref{final}) on the Fermi surface.
It is noteworthy that  emergent conformal
invariance
at criticality is
the only  way a continuous $O(d)$ rotational symmetry (which defines
a unimodular Euclidean metric) can emerge in crystalline condensed
matter, when it is not in a low-density limit.
This will produce the previously-absent physical metric in $k$-space
that is needed to define a Laplacian on the Fermi surface that is
required in a $k$-space diffusion equation.

To further emphasize the importance of the emergence of a
previously-absent metric at
criticality,  it could be speculated that the relevant critical fluctuations
of the order parameter have a description in terms of the fluctuations
of the metric, with an interpretation as
analogs of thermalized
``gravitational waves''.   Such a correspondance could provide a
concrete link
to  currently-popular  (but controversial) attempts to link high-temperature
superconductors to gravitational physics (see \textit{e.g.}, Refs.\cite{Davison,Hartnoll}).

In summary, if strong inelastic dissipation  at low temperatures in a
``strange metal'' (that
perhaps is close
to a quantum critical point) is indeed limited by attaining a  proposed
``Planckian lower  bound''\cite{Zaanen,Bruin}
$\tau$ $\ge$  $\hbar/k_BT$ on inelastic relaxation times
(\textit{i.e.}, an upper
bound  $k_BT$ on the ``quasiparticle lifetime broadening'' $\hbar/\tau$), as proposed
by Legros \textit{et al}\cite{legros}, the likely
universal formula giving the ``Planckian'' (minimum metallic) conductivity tensor
(in the absence of any Hall effect)  purely in terms of Fermi-surface
parameters is given above by equation (\ref{final}).

I acknowledge funding from the Princeton Center for Complex Materials,
a MRSEC supported by NSF Grant DMR 1420541.

\end{document}